\documentclass[twocolumn,showpacs,preprintnumbers,amsmath,amssymb,superscriptaddress]{revtex4}

\usepackage{graphicx}
\usepackage{dcolumn}
\usepackage{bm}

\def\bm#1{\boldsymbol{#1}}

\begin{document}

\preprint{Version \today}

\title{ Axion-Like Particles and Recent Observations of the Cosmic
  Infrared Background Radiation }

\author{Kazunori Kohri}%
\affiliation{
Theory Center, KEK, and the Graduate University for Advanced Studies
(Sokendai), 1-1 Oho, Tsukuba 305-0801, Japan
}%
\affiliation{
Rudolf Peierls Centre for Theoretical Physics, The University of Oxford, 1 Keble Road, Oxford OX1 3NP, UK
}%

\author{Hideo Kodama}
\affiliation{%
Yukawa Institute of Theoretical Physics (YITP), Kyoto University,  Kyoto 606-8317,  Japan
}%

\date{\today}

\begin{abstract}
  The CIBER collaboration released their first observational data of
  the Cosmic IR background (CIB) radiation, which has significant
  excesses at around the wavelength $\sim$ 1 $\mu$m compared to
  theoretically-inferred values. The amount of the CIB radiation has a
  significant influence on the opaqueness of the Universe for TeV
  gamma-rays emitted from distant sources such as AGNs.  With the
  value of CIB radiation reported by the CIBER experiment, through the
  reaction of such TeV gamma-rays with the CIB photons, the TeV
  gamma-rays should be significantly attenuated during propagation,
  which would lead to energy spectra in disagreement with current
  observations of TeV gamma ray sources. In this article, we discuss a
  possible resolution of this tension between the TeV gamma-ray
  observations and the CIB data in terms of axion [or Axion-Like
  Particles (ALPs)] that may increase the transparency of the Universe
  by the anomaly-induced photon-axion mixing.  We find a region in the
  parameter space of the axion mass,
  $m_a \sim 7 \times 10^{-10} - 5 \times 10^{-8}$~eV, and the
  axion-photon coupling constant,
  $1.5 \times 10^{-11} {\rm GeV}^{-1} \lesssim g_{a\gamma} \lesssim
  8.8 \times 10^{-10} {\rm GeV}^{-1}$ that solves this problem.
\end{abstract}

\pacs{Valid PACS appear here}
\maketitle

\section{Introduction}

The intergalactic space is not transparent for high energy photons due to the $e^- e^+$ producing reaction with background photons. This reaction occurs when the energies of two photons, $E_1$ and $E_2$, satisfy the condition $ E_1 E_2> m_e^2$ where $m_e$ is the electron mass. Hence, in particular, the opacity of the Universe for the TeV gamma-rays emitted from distant sources such as Active Galactic Nucleus (AGN, or blazar) is quite sensitive to the amount of IR background radiation with wavelength around 1 $\mu$m. This opacity can be determined by measuring the energy spectra of gamma ray sources by such as Fermi or H.E.S.S. and comparing them with theoretical predictions. Most analyses of this sort such as \cite{Franceschini:2008tp} suggested that the deduced opacity can be explained by IR photons emitted from galaxies.

Recently, the CIBER experiment released the result of their
measurements of the cosmic IR background (CIB)
radiation~\cite{Matsuura:2017lub}, which is the first direct measurement
of the diffuse background spectrum at 0.8 -- 1.7 $\mu$m. Remarkably,
they found that there exist clear excesses in their data~\footnote{For
  origins of the excuses,
  see~\cite{Santos:2002hd,Salvaterra:2002rg,Fernandez:2005gx,Mii:2005as,Yue:2012dd,Yue:2013hya}
  as relics of redshifted ultraviolet photons emitted at an epoch of
  cosmic reionization induced by Pop III stars, young galaxies, or
  black holes. For a decaying axion into the IR photon with its
  mass$\sim {\cal O}$~(1)eV, see~\cite{Gong:2015hke}. } compared with
the previous prediction theoretically-deduced by indirect
observations~\cite{Franceschini:2008tp} dubbed ``Franceschini'' or
``Franceschini-Th'' in this article.  If we adopt this CIBER result,
the observed energy spectra of the TeV gamma-rays become inconsistent
with reasonable particle acceleration models at the emission sites
because of the huge absorption by the IR background.

One possible way to resolve this inconsistency is to assume the
existence of a hypothetical particle called axion.~\footnote{See also
  \cite{Essey:2010er} and references therein for a solution by
  line-of-sight cosmic-ray interactions and continuous gamma-ray
  productions. } Axion was originally introduced to solve the strong
CP problem by the Peccei-Quinn
mechanism~\cite{Peccei:1977hh,Weinberg:1977ma}. When the chiral
Peccei-Quinn U(1) symmetry breaks down spontaneously, a pseudo scalar
particle called QCD axion appears as a Nambu-Goldstone boson. One of
the most important properties of the axion is that it couples gauge
bosons through the Chern-Simons (CS) term produced by the chiral
anomaly. For example, the CS coupling to electromagnetic fields reads
$g_{a\gamma} a \bm{E}\cdot \bm{B}$, where $a$ is the axion field, and
the coupling constant $g_{a\gamma}$ has the mass dimension minus
unity. This CS coupling induces axion-photon oscillations in external
magnetic fields $B$ in the Universe. Because the interaction of axion
with matter is extremely weak, we can utilize this oscillation to
increase the transparency of the
universe~\cite{Raffelt:1996wa,Hooper:2007bq,SanchezConde:2009wu,Belikov:2010ma}.

Axion gains mass $m_a$ by non-perturbative effects. In the case of QCD
axion, $m_a$ is proportional to $g_{a\gamma}$ and as a consequence,
constrained by various astrophysical phenomena to be larger than
$10^{-6}$eV, which is too large to solve the above problem. If we go
beyond the standard model of particles, however, other types of axions
appear. In particular, in string theories, a rich variety of axions
are expected to arise by compactification~\cite{Arvanitaki:2009fg}
universally.  To include such wider classes of axions, the term
axion-like particles (ALPs) are often used. For ALPs, the mass $m_a$
and the EM coupling constant $g_{a\gamma}$ are treated as independent
parameters. For example, see~\cite{Cicoli:2012sz} for concrete models.

In this article, we discuss the possibility to resolve the above
discrepancies among the observational data in terms of
ALPs. Actually we find an allowed
parameter region in the $(m_a,g_{a\gamma})$ plane.  We use the natural
(or Heaviside-Lorentz) system of units $\hbar= c = 1$ in this article.
Hereafter, we do not distinguish two words, axion and ALP.

\section{Axion photon conversion}

Inside a jet of an AGN, protons are accelerate to high energies and produce copious $\pi^0$'s through non-elastic
$p-\gamma$ scatterings. Then, $\pi^0$ decays into the high-energy
photons. In the present article, we assume that TeV gamma rays are
mainly produced by this mechanism~%
\footnote{Contrary to the $p-p$ scattering scenario that assumes
  $n \sim 30$~cm$^{-3}$ inside the jet, we do not have to worry about
  possible effects from plasma frequencies,
  $w_{\rm pl} ~\sim 2 \times 10^{-10}~{\rm eV} \sqrt{n/30{\rm
      cm}^{-3}}$
  on the oscillation because that effect is negligible in the current
  study for which $ n \ll 30{\rm cm}^{-3}$.}.
These TeV gamma ray photons can convert to axions through the CS term $g_{a\gamma}\b,{E}\cdot\bm{B}$ under the presence of coherent magnetic fields $B$ transversal to the photon propagation direction.
 The conversion probability between photon and axion is represented as~\cite{Hooper:2007bq,Belikov:2010ma}
\begin{eqnarray}
  \label{eq:Pga}
  P_{\gamma \leftrightarrow a} = \frac{h_{\gamma}}
          {1+\left(\frac{E_{\gamma}}{E_*}\right)^{-2} }
    \sin^2\left[
         \frac{g_{a\gamma} B r}{2}
          \sqrt{1+\left(\frac{E_{\gamma}}{E_*}\right)^{-2}}
          \right],
\end{eqnarray}
where $E_\gamma$ is the photon energy, and $E_*=m_a^2 /(2g_{a\gamma}B)$. Here, $h_\gamma = 1/2$ for $\gamma \to a$ after averaging over the photon helicity, and $h_\gamma = 1$ for $a \to \gamma$~\cite{Simet:2007sa}. We do not consider the resonant conversion~\cite{Hochmuth:2007hk} whose contribution may widen the allowed axion parameter range.

From Eq.~(\ref{eq:Pga}), we see that the probability becomes sizable when the following two
conditions are satisfied: \\
1) for the photon energy $E_\gamma$,
\begin{eqnarray}
  \label{eq:Estar}
  E_{\gamma} > E_*=m_a^2 /(2g_{a\gamma}B),
\end{eqnarray}
\noindent
2) for the distance $r$ the photon propagates,
\begin{eqnarray}
  \label{eq:rHa}
  r \gtrsim r_{ha}\equiv 2/(g_{a\gamma}B).
\end{eqnarray}
In the present article, we adopt the notation
$g_{11}\equiv {g_{a\gamma}}/10^{-11} {\rm GeV}^{-1}$,
$B_{10 \mu {\rm G}} \equiv {B}/10 \mu {\rm G}$, and
$m_{a,{\rm neV}} \equiv m_a /{\rm neV}$ with 1neV = $10^{-9}$eV.
Then, we find $E_*$ in the first condition \eqref{eq:Estar} can be written
 \begin{eqnarray}
 \label{eq:Estar2}
 E_* \sim  \frac{ 10 {\rm GeV} m_{a,n{\rm eV}}^2}{g_{11} B_{10 \mu {\rm
     G}} }
 \sim  \frac{10^2 {\rm TeV} m_{a,n{\rm eV}}^2}{g_{11}
 B_{n{\rm G}} }
\sim  \frac{10 {\rm keV} m_{a,n{\rm eV}}^2}{g_{11}
 B_{10 {\rm G}} }
\end{eqnarray}
and $r_{ha}$ in the second condition \eqref{eq:rHa} can be written
 \begin{eqnarray}
   \label{eq:rHa2}
   r_{ha}\sim \frac{10 {\rm kpc}}{g_{11}B_{10 \mu {\rm G}}} \sim
 \frac{10^3 {\rm Mpc}}{g_{11}B_{ n {\rm G}}} \sim \frac{10^{-1} {\rm
     pc}}{g_{11}B_{10 {\rm G}}},
\end{eqnarray}
respectively. Here, the three sets of values shown in (\ref{eq:rHa2}) correspond to those for $r$ and $B$ at 1) blazar jets, 2) Inter-Galactic Spaces
(IGSs), and 3) the Milky-Way (MW) Galaxy, respectively.

If these two conditions are satisfied in the strong magnetic region
around the gamma-ray source, a significant fraction of the emitted
gamma-ray photons are converted to axions. Here, note that the
condition \eqref{eq:rHa} with \eqref{eq:rHa2} has a very similar
structure to the so-called ``Hillas condition''
$r \gtrsim r_{Ha} = 10 {\rm kpc} (\frac{\epsilon}{10^{19} {\rm eV}})
(\frac{B}{10\mu {\rm G}})^{-1}{q}^{-1}$
for the acceleration of cosmic ray protons to energy
$\epsilon$~\cite{Hooper:2007bq}. This condition means that a proton
can be accelerated up to $\epsilon\sim 10^{19}$~eV if the acceleration
region size is larger than the order of its Larmor radius. In fact,
when the Hillas condition is satisfied in the TeV gamma-ray emission
region, the condition \eqref{eq:rHa} is automatically satisfied,
provided that the maximal proton energy $\epsilon_{\rm max}$ is larger
than $g_{11}^{-1}10^{19}$~eV. In the present article, we assume that
TeV gamma-ray extragalactic sources are also acceleration sites of
high energy cosmic rays up to $10^{19}$~eV at most, and that the last
condition is satisfied for the value of $g_{11}$ relevant to the
present paper.

\begin{figure}[ht]
\vspace{-0. cm}
    \includegraphics[width=80mm]{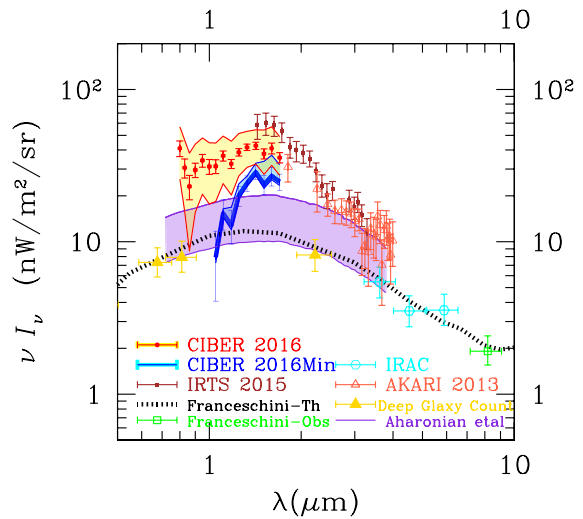}
\vspace{-1.0cm}
\caption{Observed spectra of the IR background by the CIBER
  collaboration~\cite{Matsuura:2017lub}.  Franceschini-Th means the line
  of the theoretically-inferred values reported by
  Ref.~\cite{Franceschini:2008tp}.  Hereafter we use names of the data
  to be CIBER (CIBER 2016), IRTS (IRTS 2015), and AKARI (AKARI 2013)
  in the text. We adopt a lower bound on a each bin by taking the
  lower end due to a systematic error and a 2$\sigma$ statistical
  error. For the references of these observational data,
  see~\cite{Matsuura:2017lub,Aharonian:2005gh} and references therein.  }
\label{fig:CIBwide}
\end{figure}

Axions produced around gamma-ray sources can arrive at the MW galaxy
without any absorption if they do not oscillate back into photons
during their travel in the inter-galactic space for parameters of
$m_a < n$eV and $B \lesssim {\cal O}(1)n$G (The upper bound on $B$ at
IGS). Inside the Galaxy then, axions can oscillate back into photons
if the conditions \eqref{eq:Estar} and \eqref{eq:rHa} are
satisfied. Thus, we can observe TeV gamma-rays from distant blazars
without suffering from severe absorption even with such a large amount
of CIB background radiation reported by CIBER. In
Fig.~\ref{fig:CIBwide}, we plot the data of the CIB radiations.

It is notable that the decrease of the observed isotropic IR spectrum
due to the pair-annihilation between an IR photon and a TeV gamma-ray
should be negligible compared with the total CIB radiation which is
uniformly-filled in the cosmological volume. Thus, we will not be able
to observe such a small deficit in Fig.~\ref{fig:CIBwide}.

The current model with the axion-photon conversion can solve the
tension between the observed TeV gamma-ray spectrum and the
theoretically-predicted one suffering from the significant absorption
due to the large amount of CIB radiation in the CIBER data. Thus, this
model does not modify the difference itself between the line of
Franceschini et al and the one reported by the CIBER experiment.

\section{Absorption of gamma-rays by cosmic  IR background radiation}

The threshold energy $E_{\rm th}$ of the gamma ray photon for the reaction $\gamma + \gamma_{\rm BG} \to e^+ + e^-$ is approximately given by
\begin{eqnarray}
  \label{eq:Eth}
  E_{\rm th} \sim \frac{m_e^2}{E_{\gamma_{\rm BG}}}
\sim 1 {\rm TeV}
 \left( \frac{E_{\gamma_{\rm BG}}}{1 {\rm eV}}\right)^{-1},
\end{eqnarray}
where $E_{\gamma _{\rm BG}}$ is the energy of background photons. For a
gamma-ray that was emitted at the redshift $z=z_s$ and arrives at the MW $(z=0)$ with energy $E_\gamma$ is estimated as
\begin{eqnarray}
  \label{eq:tau}
 && \tau(E_{\gamma}) =  \int^{z_s}_0 \frac{dz}{(1+z)H(z)}
 \notag\\
     &&\qquad
     \times \int^{}_{E'_\gamma\ge E_{\rm th}(E_{\gamma_{\rm BG}})}
 dE_{\gamma_{\rm BG}} \frac{dn_{\gamma_{\rm BG}}}{dE_{\gamma_{\rm BG}}}
   \bar{\sigma}(E'_{\gamma},E_{\gamma_{\rm BG}}),
\end{eqnarray}
where $\frac{dn_{\gamma_{\rm BG}}}{dE_{\gamma_{\rm BG}}} $ is the
spectrum of the background radiation at $z$ as a function of
$E_{\gamma_{\rm BG}}$, $E'_{\gamma} = (1+z)E_{\gamma}$,  and
\begin{eqnarray}
  \label{eq:barsigma}
  \bar{\sigma}(E'_{\gamma},E_{\gamma_{\rm BG}}) = \int^{1-2/E'_{\gamma}E_{\gamma_{\rm BG}}}_{-1}  d\mu
  \frac12 (1-\mu) \sigma_{\gamma \gamma \to e^+ e^-}(s).
\end{eqnarray}
Here, $\mu$ is the cosine of the angle between the photon propagation directions,
$s = 2 E'_{\gamma}E_{\gamma_{\rm BG}}
(1-\mu)$
is the Mandelstam variable for the center-of-mass energy, and the cross section  $\sigma_{\gamma \gamma \to e^+ e^-}(s)$ is given by
\begin{eqnarray}
  \label{eq:sigmaGG}
 && \sigma_{\gamma \gamma \to e^+ e^-} (s) =
  \frac{3}{16} \sigma_{\rm T}
  (1 - \beta^2)  \notag\\
&& \qquad\quad
   \times\left[
   (3-\beta^2) \ln\frac{1+\beta}{1-\beta}
   - 2 \beta (2 - \beta^2)
   \right],
\end{eqnarray}
where $\sigma_{\rm T}$ is the Thomson cross section, and $\beta = 1-4 m_e^2 /s$.

\begin{figure}[ht]
\vspace{-0. cm}
    \includegraphics[width=80mm]{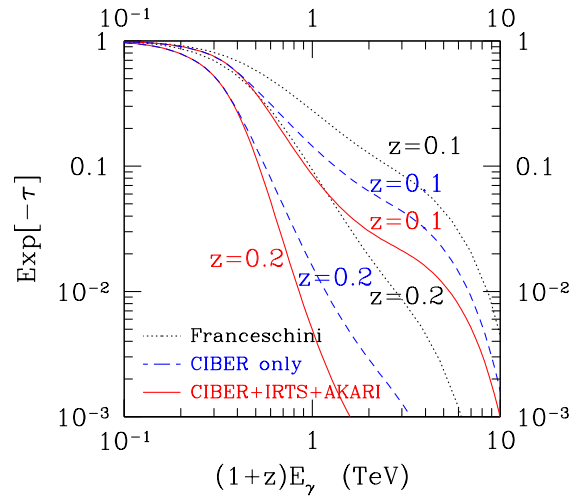}
\vspace{-1.0cm}
\caption{Absorption factors ${\rm Exp}[-\tau]$ by the IR background
  radiation for the three cases, Franceschini-Th (dotted
  line)~\cite{Franceschini:2008tp}, Franceschini-Th + CIBER (dashed
  line), and Franceschini-Th + CIBER + IRTS +
  AKARI (solid line) for two redshift values $z=0.1$ and 0.2.}
\label{fig:expTauZoom}
\end{figure}

Using these formulas, we have numerically calculated the absorption factor ${\rm Exp} [-\tau]$ of gamma-rays for three IR background radiation density models, the pure Franceschini-Th model, its combination with CIBER and with CIBER, IRTS and AKARI. Possible redshift evolutions of the IR background due to evolutions of sources are included (e.g.,
see~\cite{Franceschini:2008tp}). The results are plotted in Fig.~\ref{fig:expTauZoom},

These results can be understood in the following way. First, from the above formulas, the gamma-ray horizon length $r_{H_{\gamma}}$ can be estimated approximately as
\begin{eqnarray}
  \label{eq:rHgamma}
  r_{H_{\gamma}}(E_{\gamma}) \sim
  500 {\rm Mpc}
  \left(
  \frac{n_{\gamma_{\rm BG}}}{ 2 \times 10^{-3} {\rm cm}^{-3}}
  \right)^{-1},
\end{eqnarray}
where we have normalized $n_{\gamma_{\rm BG}}$ by the value corresponding to the IR photon density
$10 {\rm nW} {\rm m}^{-2} {\rm sr}^{-1} \sim 2 \times 10^{-3} {\rm
  eV} {\rm cm}^{-3}$
at the wavelength  $\sim 1~\mu$m or  the energy  $E_{\gamma_{\rm BG}} \sim 1.23~{\rm eV} (\lambda/\mu{\rm m})^{-1} \sim 1{\rm eV}$.
For the redshift $z$
= 0.1 ($z$ = 0.2), the distance is approximately given by
$r \sim 400$Mpc ($r \sim 700$Mpc), hence the opacity $\tau=r/r_{H\gamma}$ estimated by \eqref{eq:rHgamma} roughly reproduces the results shown in Fig.~\ref{fig:expTauZoom}.
From this figure, we clearly see that a significantly larger fraction of gamma-rays are absorbed if we include the CIB photons suggested by the CIBER data.

\section{Analyses}

\begin{figure}[ht]
\vspace{-0. cm}
    \includegraphics[width=80mm]{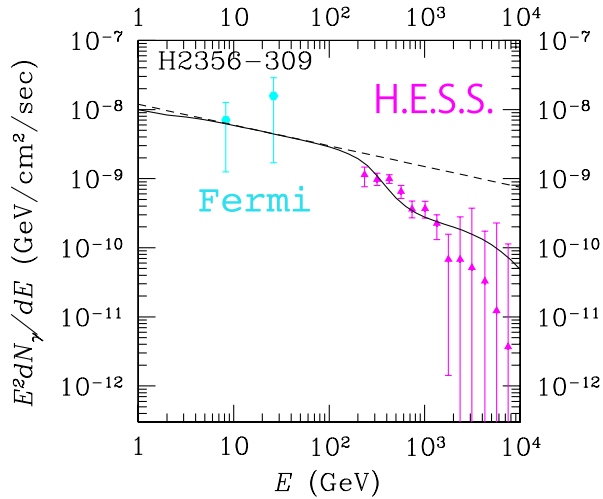}
\vspace{-1.0cm}
\caption{Gamma-ray spectrum fitted to the data of H2356 309 (the
  redshift is $z=0.165$ which gives the distance $\sim$~610 Mpc).
  Here, we adopted $g_{a\gamma}=3.2 \times 10^{-11} {\rm GeV}^{-1}$
  and $m_a=3.2 \times 10^{-9}$~eV.  The reduced $\chi^2$ is estimated
  to be $\chi^2/{\rm d.o.f} = 1.1$, which is improved from the case
  without axion $\chi^2/{\rm d.o.f} = 2.2$. The fitted value of the
  photon index is $\Gamma_s = 2.3$. We followed the way of plotting
  shown in~\cite{Belikov:2010ma}. The dashed line is the original
  spectrum at the source. }
\label{fig:H2356_309_CIBER_axion}
\end{figure}

\begin{figure}[ht]
\vspace{-0. cm}
    \includegraphics[width=80mm]{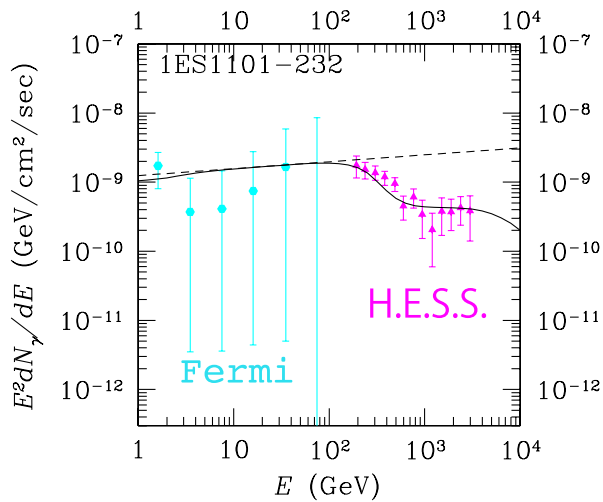}
\vspace{-1.0cm}
\caption{Same as Fig.~\ref{fig:H2356_309_CIBER_axion}, but for
  1ES1101 232 (the redshift is $z=0.186$ which gives the distance
  $\sim$ 680 Mpc.). The reduced $\chi^2$ is estimated to be
  $\chi^2/{\rm d.o.f} = 0.69$, which is improved from the case without
  axion $\chi^2/{\rm d.o.f} = 2.0$.  The fitted value of the photon
  index is $\Gamma_s = 1.9$. }
\label{fig:1ES1101_232_CIBER_axion}
\end{figure}

By choosing the axion mass $m_a$ and the axion-photon coupling
$g_{a \gamma}$, we can calculate the gamma-ray spectrum which is
observed in the MW Galaxy by tracing the oscillation between photon
and axion, and the absorption due to the reaction with the background
radiations.  In Fig.~\ref{fig:H2356_309_CIBER_axion} and
Fig.~\ref{fig:1ES1101_232_CIBER_axion}, we plot the expected spectra
obtained by this method with the source spectra being determined by
fitting to the observational data reported by Fermi~\cite{Abdo:2009}
and H.E.S.S.~\cite{Aharonian:2005gh} for H2356 309 and 1ES1101 232,
respectively. Here, we have adopted as the IR background the values of
CIBER + Franceschini-Th~\footnote{Only by adopting the data of
  Franceschini-Th, goodness of fit measures were
  $\chi^2/{\rm d.o.f}=$~0.72, and 0.97 for H2356 309 and 1ES1101 232,
  respectively. }. About the sine function in Eq.~(\ref{eq:Pga}), if
the oscillation is rapid, i.e., the phase of the sine function becomes
larger than ${\cal O} (1)$ rad, we take the averaged value for the
square of the sine function, e.g., according to
Refs.~\cite{Mirizzi:2006zy,Hooper:2007bq,Simet:2007sa,Belikov:2010ma}. This
kind of treatments can be justified because spatial fluctuations of
magnetic fields could be expected to be larger than ${\cal O} (1)$
along the line of sight toward each gamma-ray source, by which small
structures of the rapid-oscillatory features of the spectra should be
smeared out. This approach would be different from that adopted by the
Fermi-LAT collaboration in \cite{TheFermi-LAT:2016zue} for a different
target. In this paper, they took seriously precise oscillatory
features of gamma-ray spectra produced by the oscillation to axions,
although the result of their model is highly dependent on the
configuration of magnetic fields they adopted including their
coherentness.

\begin{figure}[ht]
\vspace{-0. cm}
    \includegraphics[width=80mm]{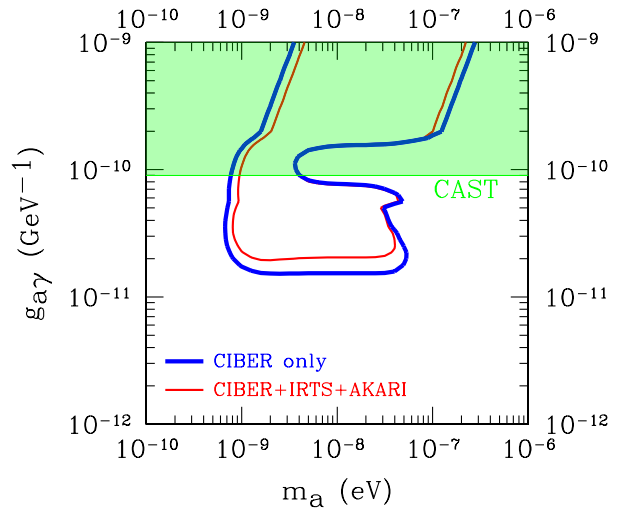}
\vspace{-1. cm}
\caption{Contours of the allowed regions in the $(m_a, g_{a\gamma})$
  plane at 95\% C.L. by the $\chi^2$ fittings with combining the
  data of H2356 309 and 1ES1101 232 for the data of CIBER +
  Franceschini-Th (outer line) and CIBER + IRTS + AKARI +
  Franceschini-Th (inner line).  Inside the contour, the tension
  between the IR background and the gamma-ray observations can be
  solved. The horizontal line gives the upper bound on $g_{a\gamma}$
  obtained by the CAST experiment~\cite{Andriamonje:2007ew}. }
\label{fig:magag}
\end{figure}

In Fig.~\ref{fig:magag}, we plot the contours of the allowed regions
in the $(m_a, g_{a\gamma})$ plane at 95\% C.L. by the $\chi^2$
fittings with combining the data of H2356 309 and 1ES1101 232 for the
data of CIBER + Franceschini-Th, and CIBER + IRTS + AKARI +
Franceschini-Th.  Inside the contour, the tension between the IR
background and the gamma-ray observations can be solved. The CAST
experiment gives the upper bound on the photon-ALP coupling to be
$g_{a\gamma} < 8.8 \times 10^{-11} {\rm GeV}^{-1}$ at 95\% C.L. for
$m_a \lesssim 0.02$~eV~\cite{Andriamonje:2007ew}. The shape of the
allowed region can be simply understood as follows. The bottom region
is excluded by the Hillas condition where Eq.~(\ref{eq:rHa}) is not
satisfied in the MW Galaxy. The right region is excluded by the first
condition shown in Eq.~(\ref{eq:Estar}) for a insufficient oscillation
into gamma-rays from axions inside the MW and the source. The left
region is excluded because of oscillations from axion into photon in
the IGS.

\section{Conclusion and Discussion}

In this article, we have discussed the problem related with the
tension between the TeV gamma-rays emitted from distant AGNs
($z \gtrsim 0.1$) and the IR background radiation which is
recently-observed by the CIBER rocket experiment
collaboration~\cite{Matsuura:2017lub}. Such a TeV gamma-ray can be
absorbed by the IR background radiation during its travel to the
Milky-Way Galaxy through the inter-galactic space. In order to solve this
problem, we have considered a possible mixing between photon and
axion. TeV gamma-rays produced in the jet of an AGN can be
converted into axions in the same site. Then the axions come to the
Milky-Way Galaxy and oscillate back to the gamma-rays there. Thus,
gamma-rays can evade absorption by the reaction with the IR background radiation.

We have found an allowed region in the $(m_a, g_{a\gamma})$ plane
under the optimistic but realistic assumptions that the sites for the
following three processes coincide; the acceleration of high-energy
protons, the production of TeV gamma-rays from those protons, and the
conversion of gamma-rays to axions~%
\footnote{For a complementary approach about 1ES1101 232, in which the
  conversion does not occur inside the jet, but inside the surrounding
  cluster of galaxy, see Ref.~\cite{Horns:2012kw}. In addition, for
  detailed analyses about possible conversion inside the jets of
  various types for blazars, see Ref.~\cite{Tavecchio:2014yoa}.}.
In this case, we obtain
$m_a \sim  7 \times 10^{-10} - 5 \times 10^{-8}$~eV for the axion mass,
and
$1.5 \times 10^{-11} {\rm GeV}^{-1} \lesssim g_{a\gamma} \lesssim 8.8
\times 10^{-10} {\rm GeV}^{-1}$
for the axion-photon coupling. This parameter region will be measured
by future experiments, ALPS II~\cite{Bahre:2013ywa},
IAXO~\cite{Armengaud:2014gea}, or
ABRACADABRA~\cite{Kahn:2016aff}. Concretely, ALPS II and IAXO are
expected to exclude $g_{a \gamma}$ down to
$2 \times 10^{-11}$~GeV$^{-1}$~\cite{Bahre:2013ywa} and
$6 \times 10^{-12}$~GeV$^{-1}$~\cite{Irastorza:2013dav},
respectively. In addition, it has been reported that ABRACADABRA will
potentially exclude much larger regions~\cite{Kahn:2016aff}.

The above assumptions were chosen to manage to solve
the problem. However, they are really reasonable because they come
from two conditions Eqs.~\eqref{eq:Estar} and
\eqref{eq:rHa}. The latter actually has a close relation  with the Hillas
condition. That means TeV gamma-rays necessarily oscillate into
axions in the source where  high-energy protons producing these TeV
gamma-rays are accelerated.

A part of the allowed region obtained in the current work might
have been already excluded by the analysis done by the Fermi-LAT
collaboration based on their observations of gamma-rays from a
different object, NGC1275~\cite{TheFermi-LAT:2016zue} (See also
excluded regions by H.E.S.S.~\cite{Abramowski:2013oea}). However, it
should be noted that the analysis in those papers strongly depends
on their models of magnetic field structures, and as a consequence
does not give the most conservative limit in the
$(m_a, g_{a\gamma})$ plane. They also used oscillatory features
of the photon-axion conversion, which is different from our method (see the
text). For similar works on the resolution of the tension with
model-dependent analyses for magnetic field structures, see also
Ref.~\cite{Meyer:2013pny} and references therein. We do not insist our
method is the best one, but we have just proposed an example of
optimistic resolution to the tension by adopting the ALPs-photon
conversion.

In the future, we expect that the Cherenkov Telescope Array will give
more precise data on TeV gamma-rays from high-redshifted
objects~\cite{Kartavtsev:2016doq}. Then the uncertainties related with
the tension between the TeV gamma-rays and the IR background radiation
will be clarified, and at the same time, information on the magnetic
field structures will be obtained.  Those attempts will open a window
to probe fundamental physics, such as the compactification of string
theory, by astrophysical phenomena.

\section*{ACKNOWLEDGMENTS}
We thank Kunihito Ioka, Shuji Matsuura, Marco Roncadelli, and Pasquale
D. Serpico for valuable discussions.  This work was partially
supported by JSPS KAKENHI Grant No.~26247042 (H.K. and K.K.) and
JP17H01131 (K.K. and H.K.), and MEXT KAKENHI Grant Nos.~JP15H05889,
and JP16H0877 (K.K.).



\end{document}